\documentclass[%
 aip,
rsi,%
 amsmath,amssymb,
 reprint,%
]{revtex4-1}

\usepackage{graphicx}
\usepackage{dcolumn}
\usepackage{bm}
\usepackage{amsmath}
\usepackage{float}
\usepackage[left=1.7cm, right=1.5cm, top=1.8cm, bottom=2.0cm]{geometry}

\begin{document}

\title[]{Inferring the dynamics of oscillatory systems using recurrent neural networks}

\author{Rok Cestnik}
\affiliation{Department of Physics and Astronomy, University of Potsdam, Karl-Liebknecht-Str. 24/25, 14476, Potsdam-Golm, Germany}
\affiliation{Institute for Brain and Behavior Amsterdam \& Amsterdam Movement Sciences, Faculty of Behavioural and Movement Sciences, Vrije Universiteit Amsterdam, van der Boechorststraat 9, 1081BT, Amsterdam, The Netherlands}
\email{rokcestn@uni-potsdam.de, r.cestnik@vu.nl}
\author{Markus Abel}
\affiliation{Ambrosys GmbH, David-Gilly-Str. 1, 14469 Potsdam, Germany}
\affiliation{Department of Physics and Astronomy, University of Potsdam, Karl-Liebknecht-Str. 24/25, 14476, Potsdam-Golm, Germany}

\date{\today}

\begin{abstract}
We investigate the predictive power of recurrent neural networks for oscillatory systems not only on the attractor, but in its vicinity as well. 
For this we consider systems perturbed by an external force. 
This allows us to not merely predict the time evolution of the system, but also study its dynamical properties, such as bifurcations, dynamical response curves, characteristic exponents etc. 
It is shown that they can be effectively estimated even in some regions of the state space where no input data were given. 
We consider several different oscillatory examples, including self-sustained, excitatory, time-delay and chaotic systems. 
Furthermore, with a statistical analysis we assess the amount of training data required for effective inference for two common recurrent neural network cells, 
the long short-term memory and the gated recurrent unit. 
\end{abstract}


\maketitle

\begin{quotation}
Inference of most dynamical properties of any system is typically best done with an active experiment, meaning that one has the power to repeatedly manipulate the system state in controlled conditions in order to isolate the desired measure, e.g. car technical inspection, where the examiners manipulate the car into specific situations in order to asses its safety on the road. Often however, one only has access to passive observations. This can be due to a number of reasons, for example, the system can be very big, like when studying the dynamics of our planet, or the system can be delicate, like when studying the phisiology of the human body. The algorithms that attempt to distil dynamical measures from passive observations commonly make assumption on how the observations were collected and typically require very long observations. Imagine, for example, assessing the car`s capabilities only from observing routine trips to work. Here we propose a conceptually simple scheme, relying on the now well established artificial neural networks. In particular, we use recurrent neural networks which have established themselves in timeseries forecasting, text generation etc. and train the networks to mimic the system dynamics, allowing us to then perform an active experiment on the trained model. We test this on several oscillatory systems and measure their characteristic properties, such as, bifurcations, dynamical response curves, characteristic exponents and compare them to the measures form the original system. 
\end{quotation}

\section{\label{sec:level1}Introduction}
Oscillatory systems can be found in all fields of natural science: in  optics~\cite{optics_book}, electronics~\cite{electronics_book}, chemistry~\cite{chemistry_book}, biology~\cite{Winfree-80}, climatology, life science, etc. 
Oscillations are present at all scales, both temporal and spatial, e.g. in biology, from cells like neurons~\cite{neurons_as_oscillators}, to organs like the heart~\cite{heart_oscillator}, to oscillations spanning the entire organism such as the circadian rhythm~\cite{circadian_book} and the menstrual cycle. 

Classical modeling of dynamical systems consists of reasoning of the terms involved followed by directly assessing the validity of the model. 
For low-dimensional systems, this works well; however the aforementioned examples are all complex, high-dimensional and coupled to their surrounding, like the brain which consists of many coupled neurons~\cite{rhythms_of_the_brain_book}. Other examples include climate models~\cite{dijkstra_2013} and fluid dynamics, which have been a major driving force for the investigation of periodic motion, synchronization of oscillatory systems~\cite{Pikovsky-Rosenblum-Kurths-01}, period doubling bifurcations, and chaotic oscillations. Accurate  modeling of such systems is hard, but with increasing computer power existing methods to infer dynamical systems from measurements are easier to realize~\cite{PhysRevLett.83.3422,PhysRevE.57.2820,IJBC04}. 

For high-dimensional systems, one has to either measure with many channels or apply embedding methods~\cite{Whitney36,Takens}, or commonly both. 
If a system is truly periodic, then it lives on a one-dimensional manifold and may be in principle modeled by a two-dimensional system of equations. 
If weakly perturbed, under certain assumptions~\cite{guckenheimer2002nonlinear}, the system remains close to the unperturbed orbit. 
Such perturbations may originate  from another oscillator, a network of oscillators or elsewhere from the environment. However, if the system is close to a bifurcation, perturbations may cause it to undergo dramatic changes in its dynamics. Bifurcations however, are hard to predict for heuristic models, whereas this is generally easier if equations are known. Under this point of view previous approaches using symbolic regression methods~\cite{PhysRevE.57.2820,PhysRevE.94.012214,lipsonschmidt09} proved successful. Heuristic methods such as liquid state machines, echo state networks, or various types of artificial neural networks~\cite{Ott_chaos_from_data,Ott_Lyapunov,Ott_attractor_reconstruction,Zimmermann-Parlitz-18,Goodfellow-et-al-2016} perform very well in predicting dynamical systems. However, few studies are known for particular aspects of oscillatory systems inferred from time series. Here, we investigate several oscillatory models under perturbation, as they may occur in real measurements. Our focus is on the inference of dynamical properties, bifurcation behavior and chaos, even if not all of the parameter variation is included in the measurement.

In all of the above methods one typically \textit{a priori} assumes a model (or a class of models), sets an optimization criterion (e.g. least squares) and optimizes model parameters or its functional constituents for nonparametric methods. Mathematical aspects are most often left aside, e.g. basic assumptions on the existence of solutions and robustness under perturbation, in particular for heuristic methods. 
Here we utilize the widely used artificial neural networks (ANNs). 
An ANN has several hyperparameters such as the actual topology of the network, the activation function, the learning rate,  and is in general very pliable toward many different tasks. Since we consider time series, we investigate the capacities of recurrent neural networks~\cite{RNN_review} (RNN). Due to loops in their connectivity they retain past information, i.e.  they inherently possess memory, similar to embedding. 
They tend to be particularly successful in speech recognition~\cite{speach_recognition}, text generation~\cite{RNN_text_generation} and machine translation~\cite{RNN_machine_translation}, where a forward-oriented semantic is present. 
The aim of this study is to evaluate how suited RNNs are for modeling oscillatory systems under the aspect of parameter change and perturbations. 
In this way, the inferred model of the oscillator can be probed via changing the perturbation signal, effectively allowing the performance of an active experiment. 

The article is structured as follows: in section~\ref{sec:previous_work} we refer to relevant related works and briefly recall the RNN functioning. 
in section~\ref{sec:inference_scheme} we introduce the dynamical inference setup and 
training scheme. We then present numerical tests with example systems in section~\ref{sec:numerical_tests} 
where we compare signal reconstruction and other observables such as the phase response curve~\cite{PRC_scholarpedia} (PRC) and the maximum Lyapunov exponent~\cite{Lyapunov_scholarpedia}. 
The different example systems are chosen as representatives of different mechanisms giving rise to oscillatory behavior; specifically, self-sustained and excitatory oscillations, time-delay induced oscillations and chaos.
We continue by presenting numerical tests on data requirement for successful inference in section~\ref{sec:data_req} 
where we compare the inference quality for different lengths of time-series used for training. 
We present the methods used in more detail in section~\ref{sec:methods} and finally, discuss the novelties, 
limitations and generalizations of our approach in the discussion section ~\ref{sec:discussion}.

\subsection{\label{sec:previous_work}Previous work}
In this paragraph we chronologically go through works related to this paper. 
In Ref.~\cite{Pearlmutter_1989} the author uses a RNN for learning state space trajectories. 
In Ref.~\cite{Funahashi_1993} the authors show that any trajectory generated by a finite-dimensional dynamical system can be effectively represented with a neural network.  
In Ref.~\cite{Withers_1998} the authors model a dynamical system with a perturbation using a RNN. 
In Ref.~\cite{FNN_biffs_2015} the authors use 
feed-forward neural networks~\cite{FNN_review} to model dynamical systems. 
They feed in delayed values of one variable as well as a control parameter as inputs and train the network for one step predictions. 
The approach works well and they reproduce bifurcation diagrams of several example dynamical systems.  
In our approach we train RNNs for one step prediction where the input consists of several time-delayed values of one or more variables 
as well as an arbitrary number of perturbative signals (we will refer to the perturbative signal inputs as p-inputs). 
The past values of one or more variables contain information of the topology of the attractor of the complete system according to the Takens' delay embedding theorem~\cite{Takens}. 
The RNN topology prioritizes more recent values over older ones for the next prediction, 
therefore, we believe it is more suitable for time-series prediction and demonstrate its efficiently throughout this paper. 

\subsection{Recurrent neural networks}
Artificial neural networks is nowadays a relatively broad term as many different network topology classes are commonly used for dealing with different types of problems.  
The simplest class of ANNs are the feed-forward networks, they have directed connections between subsequent layers without any loops, effectively allowing the information to flow in only one direction - forward. 
The slightly more general class are the recurrent neural networks (RNN), they can have loops in their connectivity, which can result in internal state memory. 
Different RNNs then differ in the fine architecture of the basic cells, the order and type of logical operations. 
In this work we apply two commonly used cells, the long short-term memory cell~\cite{LSTM} (LSTM) and the gated recurrent unit~\cite{GRU} (GRU). 
LSTM was constructed first in an attempt to deal with long term dependencies and GRU emerged as its faster simplification. 
Further details on the functioning of different RNN cells can be found in Ref~\cite{understanding_RNN}. 
The software implementation was accomplished with the help of TensorFlow~\cite{tensorflow} and Keras~\cite{keras}.

\section{\label{sec:level1}Results}
\subsection{\label{sec:inference_scheme}Inference scheme}
Consider a general dynamical system $\dot{\vec{x}}(t) = f(\vec{x})\;\in \mathbb{R}^{N}$, perturbed by an external perturbation $\vec{p}(t)\in \mathbb{R}^{N}$. 
Suppose we have measured the timeseries of $n_x \geq 1$ state variables $\vec{x} = (x_1,\ x_2,\dots,x_{n_x})$ as well as the $n_p$ timeseries of the perturbation $\vec{p} = (p_1,\ p_2,\ p_3,...)$ over a period of time. The question we investigate is if it is possible to recover both the autonomous dynamics of the system $\vec{x}$ and the system's response to the perturbation using RNNs. 
Without perturbation we can only recover the dynamics on the attractor, but with a perturbation the phase space around the attractor is explored and we have a means to infer the neighboring phase space, too. 

We train the RNN to receive historical values of $\vec{x}(t),\vec{p}(t)$ and return the time-evolved state $\vec{x}(t+\Delta t)$. 
In practice, this is accomplished by first "unrolling" the network. 
The RNN at each time step can be represented as a separate copy of the same network, where the recurrent connections have been replaced with regular connections linking every copy with its successor. 
Then this chain of networks is truncated and a finite number of "rolls" (network copies) considered.  
The historical values effectively correspond to a time-delay embedding, allowing the RNN to infer the state of the system. 
The number of rolls therefore corresponds to the dimensionality of the time-delay embedding, although the time steps we use are typically much smaller then delays used in embeddings, making the succesive steps considerably correlated. 
Nevertheless we can deduce from the Takens' embedding theorem~\cite{Takens} that at least $R > 2M+1$ rolls have to be considered, where $M$ is the dimensionality of the attractor. 
The sampling $\Delta t$ must be smaller than the smallest time scale which occurs in the system (or which we may want to include in our modeling). Heuristically, one can say that the time-resolution $\Delta t$ should be chosen fine enough to see the details of interest. 

Given an appropriate resolution and number of network copies we can begin to "train" our model, i.e. to start a loop for the statistical inference method: 
At each training step the network "learns" the possible relation:
\begin{equation}
 \vec{x}(t),\vec{p}(t) \mapsto \vec{x}(t+\Delta t)
\end{equation} 
using the time instants 
$t,\ t - \Delta t,\ t - 2\Delta t,...,\ t - (R-1)\Delta t$. We use a least-squares optimization criterion $ \frac{\lVert \vec{x}_\text{e} - \vec{x} \,\, \rVert \hfill}{\lVert \vec{x} - \langle \vec{x} \rangle  \rVert \hfill} $ 
(where $\langle \cdot \rangle$ stands for the mean) to determine quantitatively how well the estimates $\vec{x}_\text{e}$ match the true values $\vec{x}$. Hereafter, we use estimated and modeled as synonymous.

\subsection{\label{sec:numerical_tests}Examples}
We put our scheme to the test on several model systems, including time-delay, excitatory and chaotic oscillators. The \textit{validation test} consists of comparing the modeled signal with the original when presented with data never seen in training. As important measures of oscillatory systems we estimate the phase response curve~\cite{PRC_scholarpedia} (PRC) and the maximal Lyapunov exponent~\cite{Lyapunov_scholarpedia} for comparing the predictive power the model has in a dynamical systems context.

For all examples shown in this paper we use a network with $1$ hidden layer of $32$ nodes and $36$ rolls. 
We use {\it tanh} activation for all but the output layer, where we use {\it linear} activation so that a continuous signal can be produced. There are two common cells used in RNN: the long short-term memory cell~\cite{LSTM} (LSTM) and the gated recurrent unit~\cite{GRU} (GRU). We tested both for the systems in this study; as a result we found that GRU performed poorly, hence all results shown are for LSTM models, cf. Sec.~\ref{sec:data_req} for a comparison of the two cells. 
To generate the data we first simulate the perturbation signal using the stochastic Euler-Maruyama integration scheme, and then integrate the dynamical equations with fourth order Runge Kutta. 
We use a sufficiently small time step and then re-sample the signals to an appropriately lower time resolution to create the network training data. 
The resolution is chosen such that $36$ points (the number of considered historical values $R$) corresponds to $1$ natural period of the oscillator.
In the case of chaotic oscillators, this was computed as the average period, in the case of excitatory systems the time needed to return from the excited state to the fixed point was used. 

\subsubsection{\label{sec:level3}Roessler oscillator - phase response curve, bifurcation diagram and Lyapunov exponents}
For our first test we use the Roessler system~\cite{roessler}, because it exhibits many different regimes, i.e. simple periodic oscillations, higher period oscillations and chaos, 
by varying just one parameter $b$, cf. Fig.~\ref{fig2}(a) for the bifurcation diagram. The corresponding equations, including the perturbation $p$ read: 
\begin{equation}
\begin{split}
\dot{x} =& \ - y - z \\
\dot{y} =& \ x + ay \\
\dot{z} =& \ b + z(x-c) + p(t)
\end{split}
\label{eq:roessler}
\end{equation}
with parameters $a = 0.2$ and $c = 5.7$. To explore the phase space, we can vary $b$ through a constant term in the perturbation $p(t)$. For the first test we set $b = 2.0$, such that the system has a simple attractive periodic orbit. In the following we use a stochastic perturbation, generated by an Ornstein-Uhlenbeck~\cite{ornstein-uhlenbeck} process:
\begin{equation}
\dot{q} = -q/\tau + \epsilon \sqrt{2/\tau} \xi(t)
\label{eq:inp1}
\end{equation}
where $\xi$ is Gaussian white noise $\langle \xi(t) \xi(t') \rangle = \delta(t-t')$, and $\epsilon = 0.5$ and $\tau = 5.0$ are the amplitude and correlation time of $q$.
The stochastic differential equation is integrated with the Euler-Maruyama method to obtain noise with exponentially decaying correlation: $\langle q(t) q(t') \rangle = \epsilon^2 e^{-(t-t')/\tau}$.

Now, we set $p(t) = q(t)$ and feed both the signal $x(t)$ and the perturbation signal into the network as described in section~\ref{sec:inference_scheme}. 
The timeseries length corresponds to $1000$ natural periods, which is presented to the network during $500$ training epoch in batches of $100$ time points with resolution $\Delta t = 0.17$ (the time step used for the integration is significantly smaller). 
The network is trained using stochastic gradient descent~\cite{SGD} with learning rate $0.005$.

\begin{figure}[H]
\includegraphics[width=1.02\columnwidth]{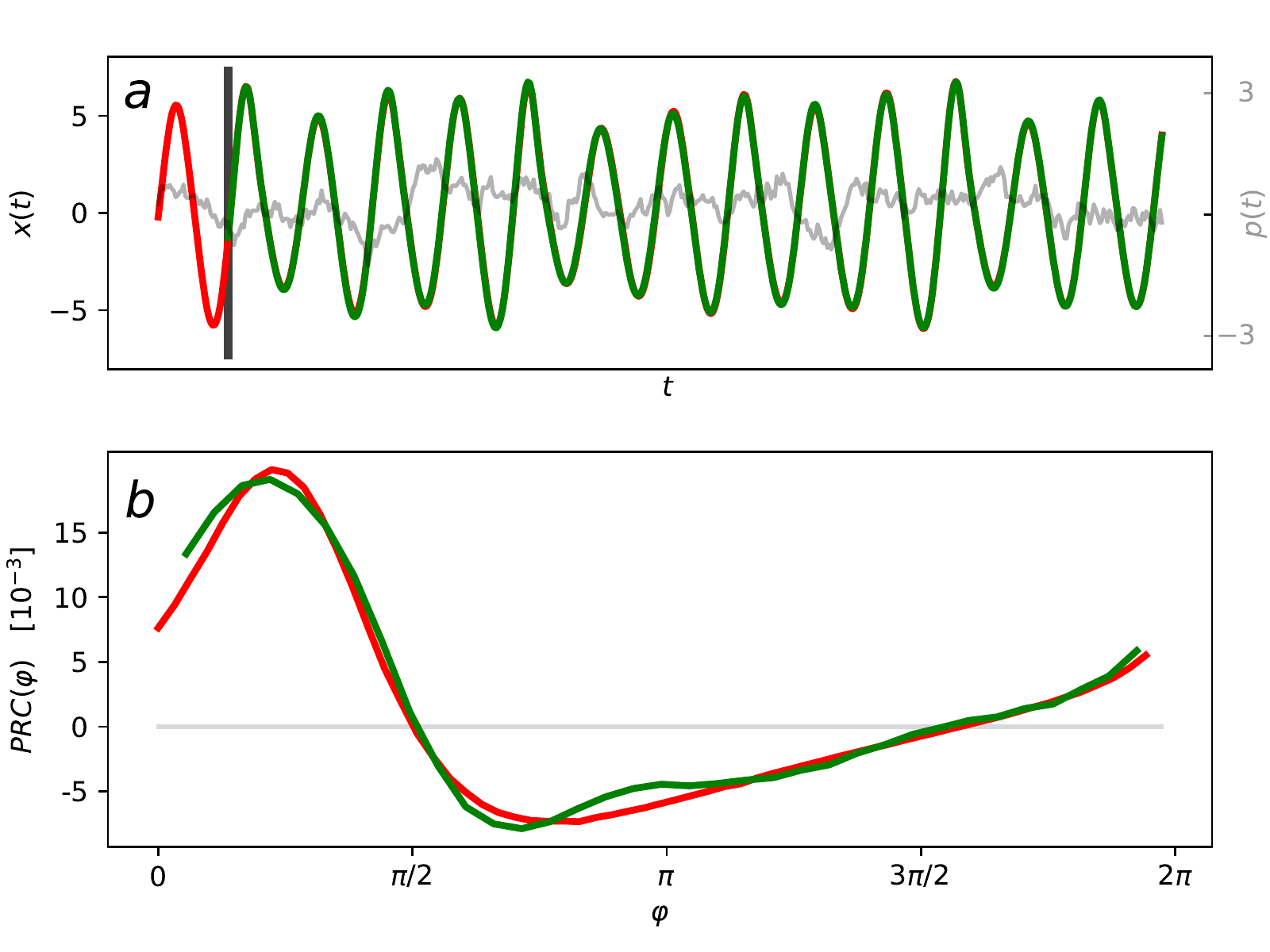}
\caption{a: The training signal generated with Eqs.~(\ref{eq:roessler}) in red and the RNN reproduced signal in green. Both have the same p-input, depicted with gray (scaled for being visually comparable to the signal). 
The vertical black line marks the beginning of the forecast. 
b: The true phase response curve of system~(\ref{eq:roessler}) in red and the one inferred from the RNN in green.} 
\label{fig1}
\end{figure}

\begin{figure}[H]
\includegraphics[width=0.97\columnwidth]{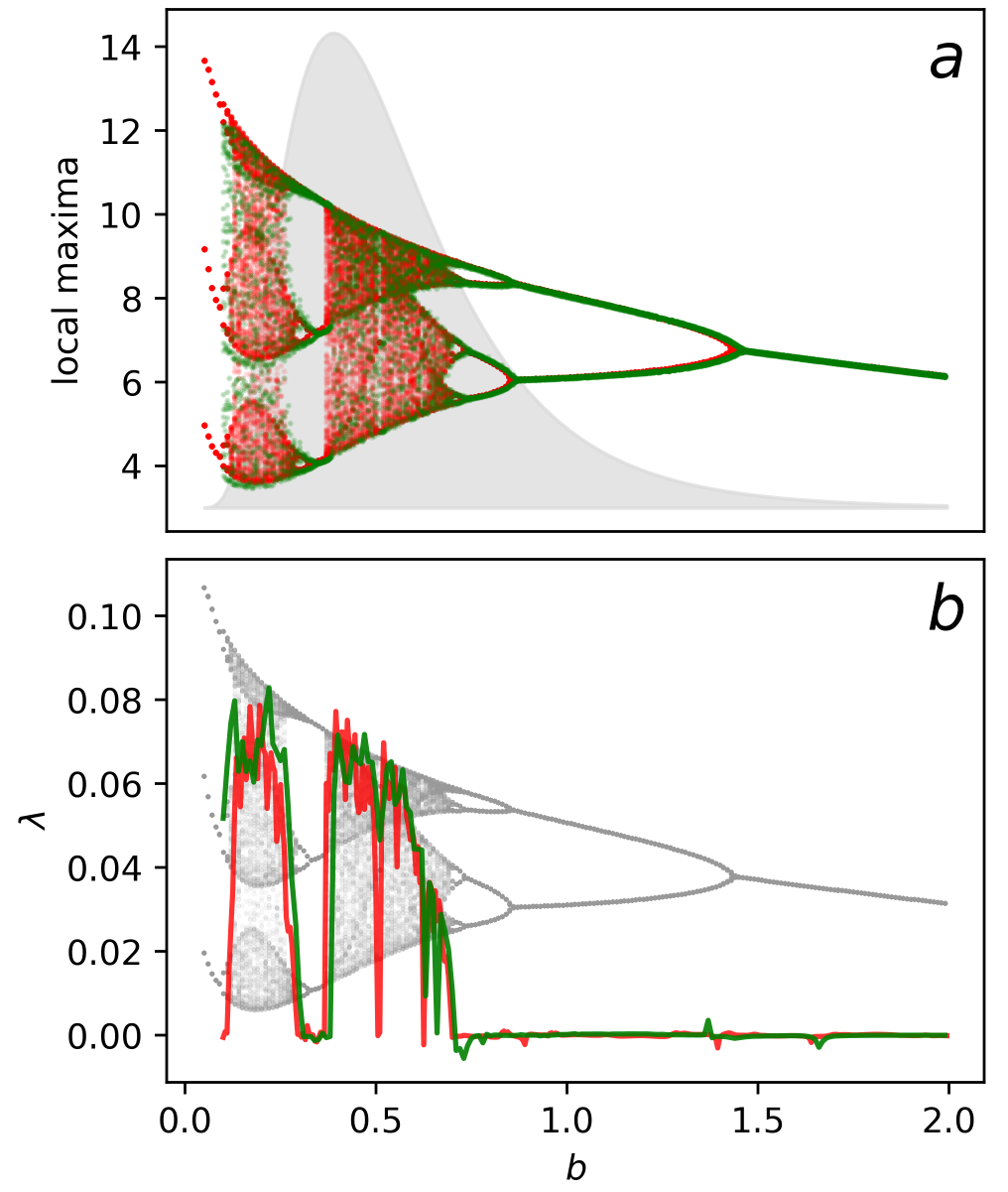}
\caption{a: The bifurcation diagram of system~(\ref{eq:roessler}) in red and the one inferred from the RNN in green. The probability density of the p-input is underlaid in gray (scaled for being visually comparable to the diagram). 
b: The Lyapunov exponents of system~(\ref{eq:roessler}) in red and the ones inferred from the RNN in green. The true bifurcation diagram is underlaid in gray (scaled for being visually comparable). 
Note that the bifurcation diagram as well as the Lyapunov exponent range were reproduced with a single RNN trained on correlated noise p-input, Eq.~(\ref{eq:enap}).}
\label{fig2}
\end{figure}

The network learns to reproduce the dynamics to a mean deviation 
of $2.5\times 10^{-2}$ (for the time window in Fig.~\ref{fig1}), such that the reproduced signal is visually indistinguishable from the one generated with Eqs.~(\ref{eq:roessler}). 
This holds true for both the perturbed signal - where $p(t)$ is fed to both the network and the equations, as well as for the unperturbed signal - where $p(t) = 0$ is used. 

Can we use the inferred network for more than just mimicking a signal, e.g. to study dynamical regimes? We want to study this scenario in probing the network for dynamical responses to stimuli. Since the system in question is a self-sustained oscillator it is natural to estimate its PRC, cf. Sec.~\ref{sec:PRC_method}. The comparison of the estimate  obtained from the RNN and the the true PRC, is displayed in Fig.~\ref{fig1}. The coincidence is very good, up to mean deviation of $0.1$ in the entire phase range $[0,2\pi)$. Indeed, this can be an effective method of inferring the PRC from data, cf.~\cite{Ota-PRL-2009, Imai-2017, Cestnik-Rosenblum-18}. 

We perform another test with the Roessler oscillator, this time testing the power of the network to reproduce the system across several dynamical regimes. 
For this we use $b = 0$ and a strong and varied p-input that considerably explores the state space: 
\begin{equation}
p(t) = \frac{1}{2} \exp(q(t))
\label{eq:enap}
\end{equation}
where $q(t)$ is the p-input described in Eq.~(\ref{eq:inp1}). It yields a process with a log-normal distribution: $P(p) \sim \frac{1}{p} \exp\Big(-2 \big(\log(p)+\log(2)\big)^2\Big)$. Such p-input spans a wide range of values, effectively introducing different regimes of our system to the network, see Fig.~\ref{fig2}(a) for the p-input probability distribution with respect to $b$ bifurcation (grey shaded region in the background).
The idea is that the network then effectively learns to mimic the regimes corresponding to different values of $b$, which we can invoke via the offset of the p-input $p(t)$. 
For this study we use a longer timeseries corresponding to $10000$ natural periods, which is presented to the network during $1000$ training epochs.

As a result, we find that the network reproduces the signal of the system perturbed by Eq.~(\ref{eq:enap}) well. 
Furthermore, we can estimate the bifurcation diagram from the network by feeding it different values of constant p-input, effectively setting the parameter $b$ of the model~(\ref{eq:roessler}) and observing the stationary signals, see Fig.~\ref{fig2}(a). 
In the value range of the p-input~(\ref{eq:enap}) the diagram obtained from the RNN matches the true one closely. 
It reproduces simple oscillatory regimes, chaotic regimes and the period doubling bifurcation.
Throughout the range of $b$ the natural frequency (average frequency in the case of chaos) matches the true one closely, with mean deviation of $5 \times 10^{-2}$. 

In chaotic regimes the maximum Lyapunov exponent~\cite{Lyapunov_scholarpedia} is an important measure as it quantifies the divergence of nearby trajectories in time. 
We estimate it from the RNN and plot it against the true values, see Fig.~\ref{fig2}(b).
This is accomplished by long time observation of the evolution of two nearby states, while re-scaling their difference to prevent them from diverging far from each other, see section~\ref{sec:Lyapunov_method} for further details. 
This can be an effective method for inferring the Lyapunov exponent from data, cf.~\cite{Wolf_lyapunov_algorithm, Rosenstein_lyapunov_algorithm}.

\begin{figure*}
\includegraphics[width=1.8\columnwidth]{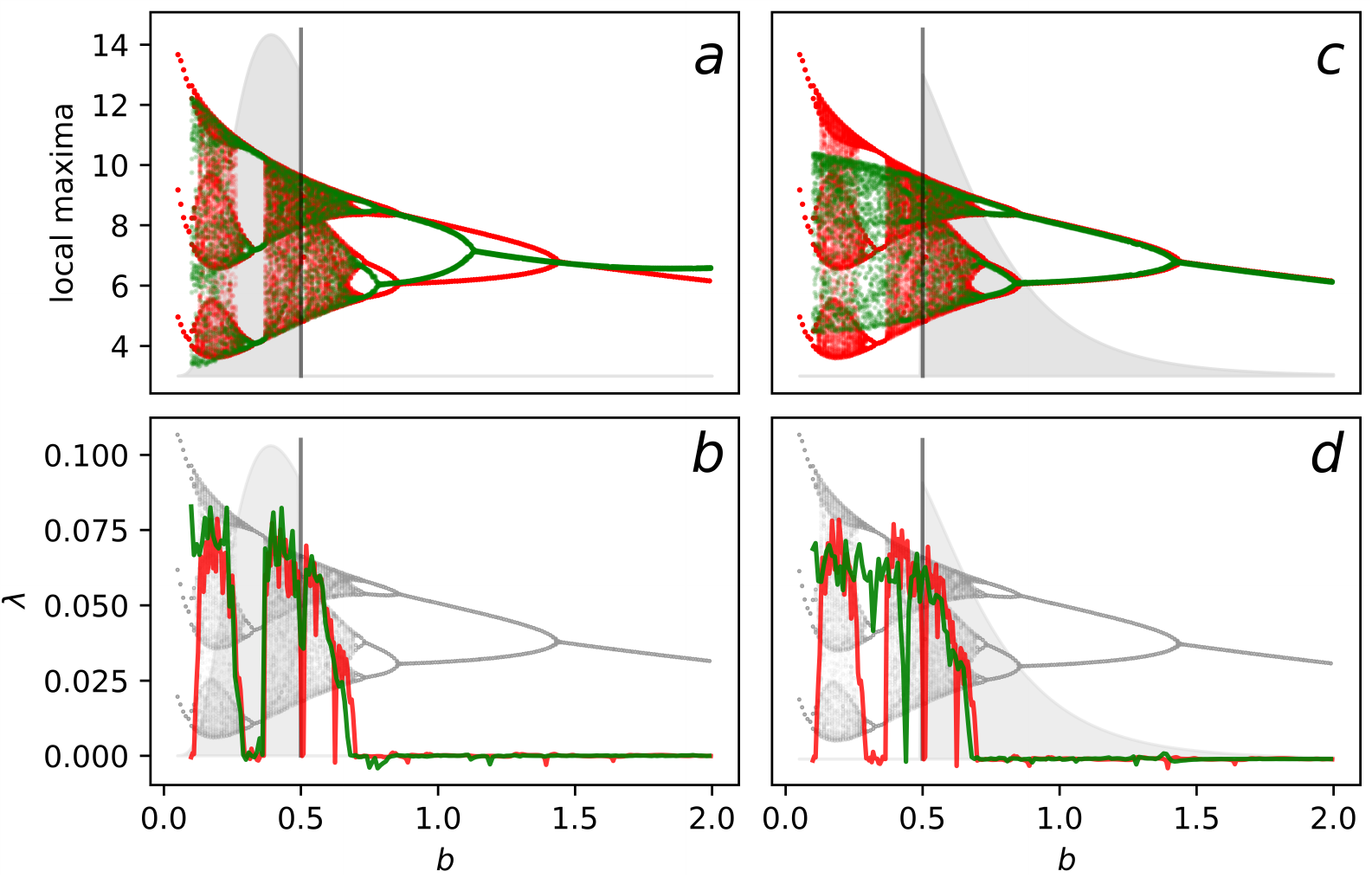}
\caption{The bifurcation diagram and maximal Lyapunov exponent for system~(\ref{eq:roessler}) perturbed by Eq.~(\ref{eq:enap_abs}). On the left (a,b) the negative exponent (-) is used, therefore limiting the p-input values below $0.5$, and on the right (c,d) the positive exponent (+) is used, limiting the p-input values to above $0.5$, see the underlaid probability distributions in gray. 
Therefore, each side has a range of parameter values $b$ that the RNN has not been presented with during training. 
In all subplots the true values of the system~(\ref{eq:roessler}) are depicted with red and the RNN inferred ones with green. }
\label{fig:halfhalfstudy}
\end{figure*}

Now we go even one step further and test the prediction of the RNN when presented with an input outside the range of trained values. 
We train two networks on slightly modified p-inputs:
\begin{equation}
p(t) = \frac{1}{2} \exp(\pm |q(t)|)
\label{eq:enap_abs}
\end{equation}
This effectively splits the probability distribution of the p-input~(\ref{eq:enap}) in two at the value $b = 0.5$. 
One network is trained only on values smaller than 0.5 and the other only on larger ones. 
Then we perform the same prediction analysis as in the previous test, estimating the bifurcation diagram and the Lyapunov exponents accross the full range of $b \in (0,2)$. 
It stands to reason that the predictions in regimes far from those presented during training will have little to do with the original system, but nevertheless it is surprising just how much can be deciphered from them. 
For example, in Fig.~\ref{fig:halfhalfstudy}(a) a period doubling bifurcation occurs outside of the trained regime, as it does in the original system (although the critical values are shifted), 
and throughout the entire test range the system remains oscillatory (it does not settle to a fixed point). 
Not all features are reflected however, for instance in Fig.~\ref{fig:halfhalfstudy}(c) in the chaotic regime outside of the p-input range the period-3 window is not observed.  

\subsubsection{\label{sec:level4}FitzHugh-Nagumo oscillator - example of an excitable system}
\begin{figure}[H]
\includegraphics[width=0.97\columnwidth]{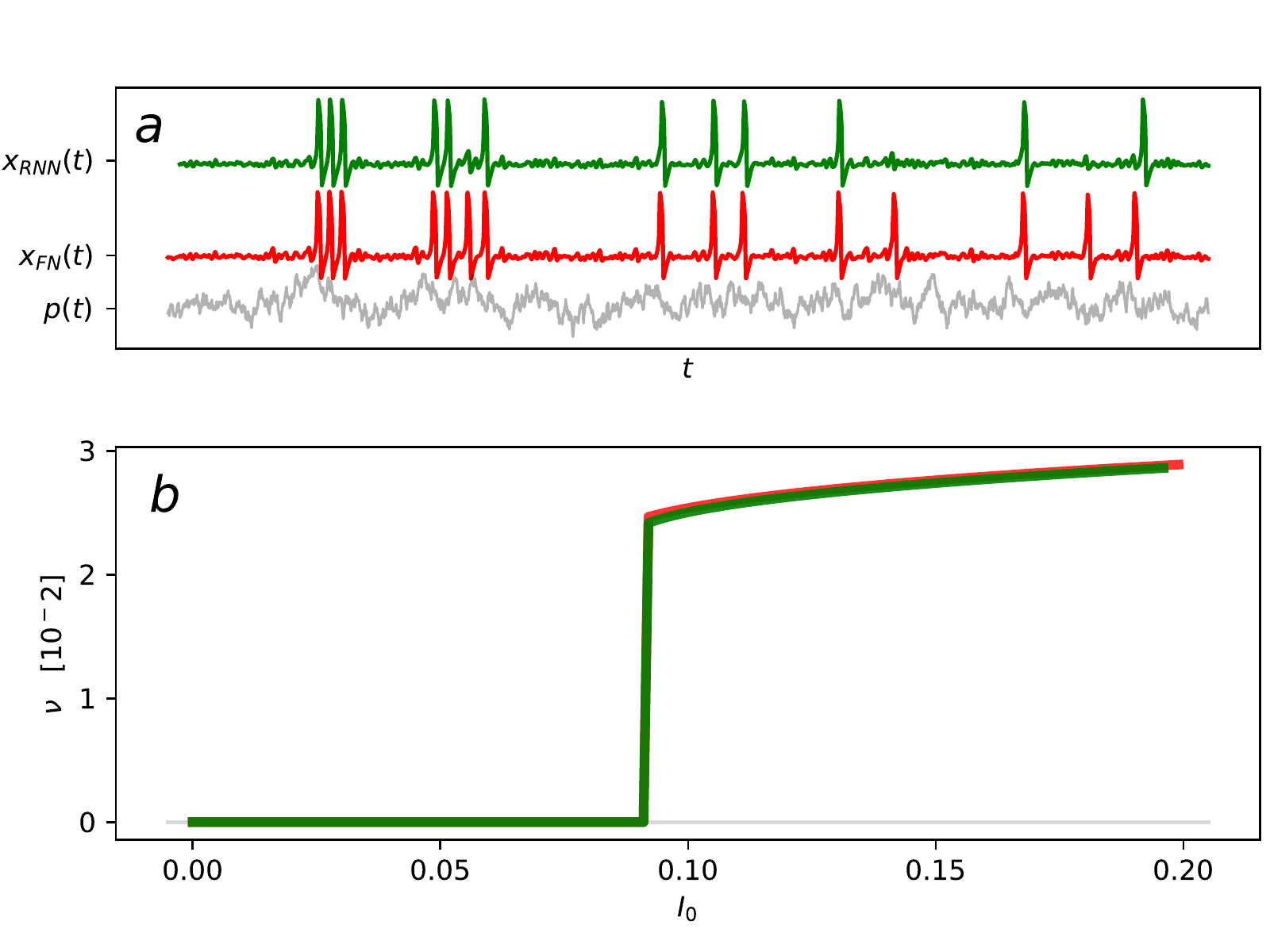}
\caption{a: The training signal generated with Eqs.~(\ref{eq:fn}) in red and the RNN reproduced signal in green (shifted up for distinction). Both have the same p-input realization, depicted with gray (scaled and shifted for being visually comparable to the signal). b: The true spiking rate of system~(\ref{eq:fn}) in red and the one inferred from the RNN in green.}
\label{fig3}
\end{figure}

In the following two examples we want to study the power of RNN for two systems with different origin and dynamical behavior of oscillations. As a first important class we investigate an excitatory system, namely the FitzHugh-Nagumo oscillator~\cite{fitzhugh_nagumo}:

\begin{equation}
\begin{split}
\dot{x} =& \ x - x^3/3 - y + I_0 + p(t)\\
\dot{y} =& \ \sigma(x + a - by)
\end{split}
\label{eq:fn}
\end{equation}
where parameters are $\sigma = 0.1$, $a = 0.7$, $b = 0.8$ and $I_0 = 0.25$. 
For the p-input we use, as once before $p(t) = q(t)$, described by Eq.~(\ref{eq:inp1}), with $\epsilon = 0.05$ and $\tau = 25.0$. The RNN is trained on timeseries comprising of $1000$ spikes, over $500$ training epochs. The time resolution is $\Delta t = 1$. 
For the excitatory oscillations we compare how well does the model reproduce a spike train when presented with a novel p-input realization, see Fig.~\ref{fig3}(a). 
Furthermore, we estimate the spiking frequency with respect to the input current $I_0$ (the p-input) and compare it to the true one, see Fig.~\ref{fig3}(b). 
For low input currents $I_0$ the system~(\ref{eq:fn}) is quiescent, i.e. it does not fire and remains close to its fixed point. 
When $I_0$ is increased, a bifurcation occurs, a limit cycle is born and the system begins to spike regularly. 
The corresponding first-order phase transition is clearly inferred from the RNN, with the critical value of the input accurately predicted up to the order $10^{-3}$. 
In addition, the estimated frequency values match the true ones closely, with mean deviation $10^{-2}$. 

\subsubsection{\label{sec:level5}Mackey-Glass equation - example of a delay system}
\begin{figure}
\includegraphics[width=1.00\columnwidth]{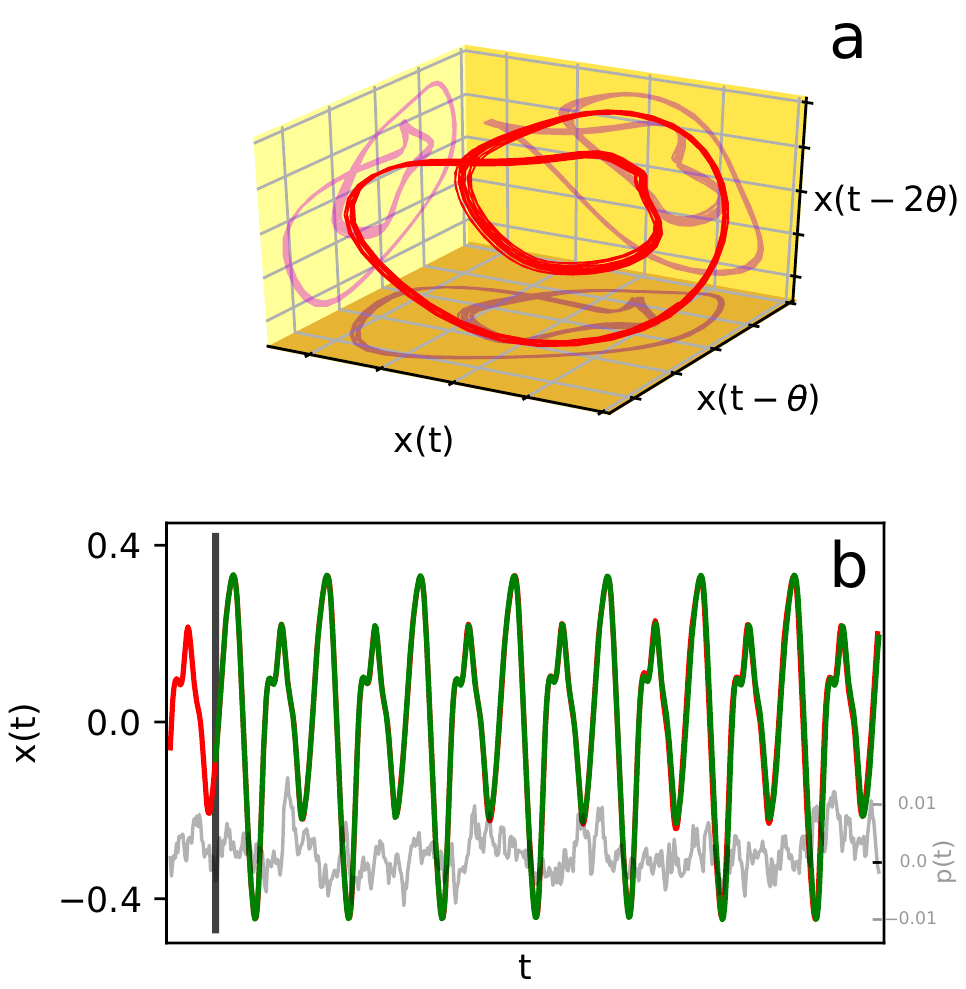}
\caption{a: Delay embedded trajectory of system~(\ref{eq:mg}) in red, its two dimensional projections in purple. For the chosen set of parameters the system has a stable limit cycle, the variation is due to the p-input. 
b: The signal generated with Eqs.~(\ref{eq:roessler}) in red and the RNN reproduced signal in green. Both have the same p-input realization, depicted with gray (scaled and shifted for being visually comparable to the signal). The vertical black line marks the beginning of the forecast. }
\label{fig4}
\end{figure}

For our final study we briefly report on the RNN results for the Mackey-Glass equation~\cite{Mackey_Glass_eq}, as a representative of time-delay systems:
\begin{equation}
\dot{x} = a \frac{x_{\theta}}{1+x_{\theta}^n} - bx + p(t)
\label{eq:mg}
\end{equation}
where $x_{\theta}$ represents the time delayed variable $x(t-\theta)$, $a = 2$, $b = 1$, $n = 8$ and the time-delay $\theta = 2$. 
In this parameter regime the equation yields a stable limit cycle with a period-2 orbit, see Fig.~\ref{fig4}. 
We use p-input $p(t) = q(t)$, Eq.~(\ref{eq:inp1}), with $\epsilon = 0.005$ and $\tau = 1.0$. 
The length of time series corresponds to $5000$ natural periods, over $500$ epochs in batches of $100$ time points with resolution $\Delta t = 0.15$. 
The dynamics is well reproduced with a mean deviation of $5 \times 10^{-2}$. 
Intuitively, a RNN seems to be suited well for modeling delay equations, since it has inherent delay. 
We conclude that for this important model class RNNs work well. 

\subsection{\label{sec:data_req}Amount of data and noise study}
\begin{figure*}
\includegraphics[width=1.6\columnwidth]{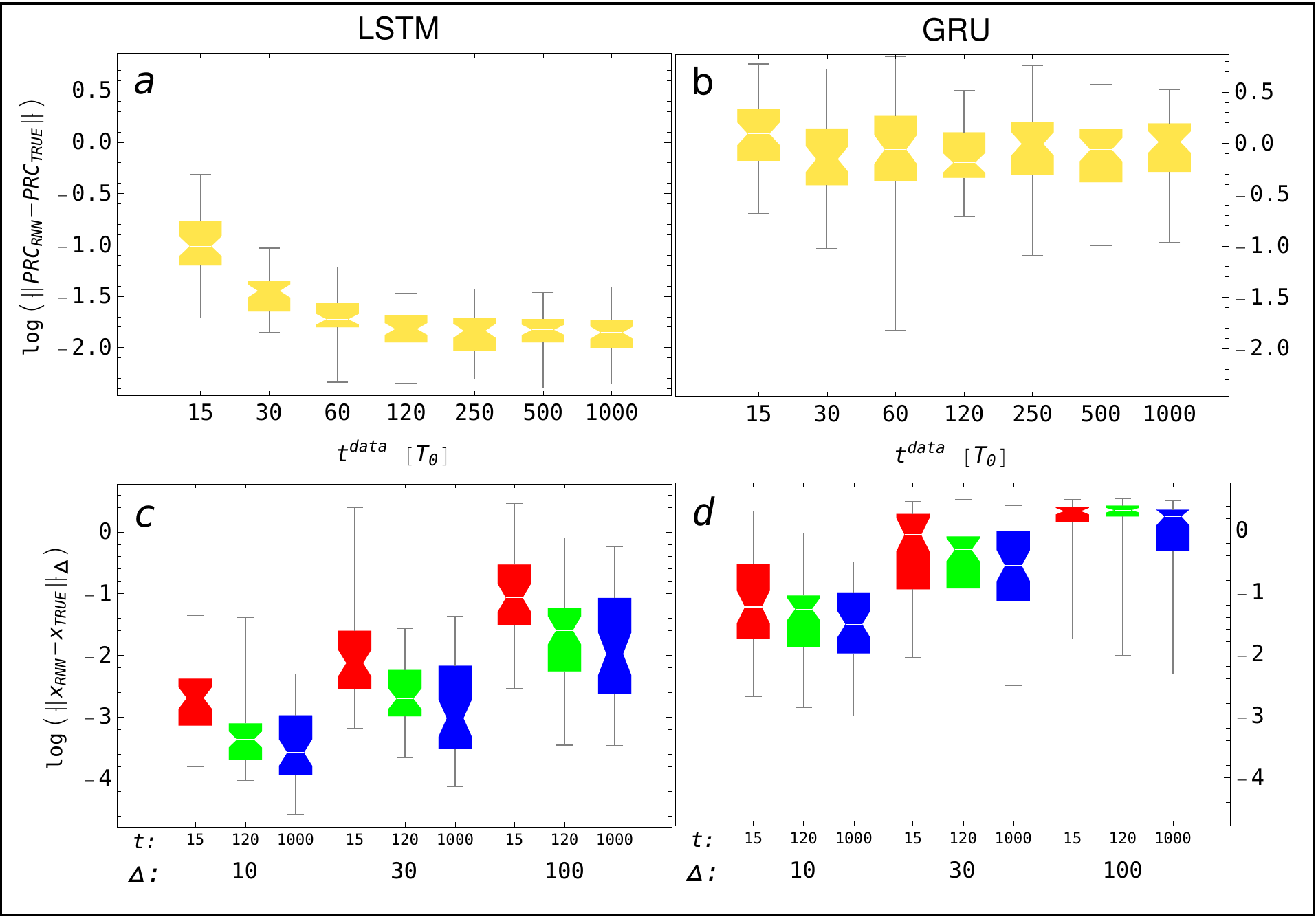}
\caption{Comparison of data requirement for two different cells, LSTM left (a,c) and GRU right (b,d). 
In the top plots (a,b), the error of the inferred PRC with respect to the length of data provided, $t^{data}$ (in units of the natural period $T_0$). 
In the bottom plots (c,d), the error of the reproduced signal with respect to the length of data provided $t^{data}$ (15, 120 and 1000) for three different forecast lengths, $\Delta$ (10, 30 and 100).
}
\label{fig:datastudy1}
\end{figure*}

Any good statistics-based study includes a section on the dependence of the result on the amount of data provided and the sensitivity to noise - we do so in the following paragraphs. 
We present only results for the Roessler oscillator, Eq.~(\ref{eq:roessler}). 
We train independent RNN models with different lengths of timeseries. We vary the amount of data supplied to the RNN in the following way: we keep the product of the timeseries length and number of epochs constant, thereby always introducing the same number of data points to the network $(500000)$, i.e. we change the number of occurrences of the same points. 
The sampling rate is kept constant. 
We test the range from $15$ to $1000$ periods and measure the error of the PRC and signal, see Fig.~\ref{fig:datastudy1}. For each set of parameters 100 models are trained and evaluated. 
The PRC error is evaluated as the $L_2$ norm of the difference between the true and the reconstructed curve: 
$$\int\limits_0^{2\pi} \Big( PRC_{_{RNN}}(\varphi)-PRC_{_{TRUE}}(\varphi) \Big)^2 d\varphi$$
and similarly for the signal error:
$$\int\limits_0^{\Delta} \Big( x_{_{RNN}}(t)-x_{_{TRUE}}(t) \Big)^2 dt$$
where we further have to determine over what interval we evaluate it, $\Delta$. 

The results of this evaluation are shown in Fig.~\ref{fig:datastudy1}. Here we also demonstrate the difference between LSTM and GRU cell types, 
underlining our previous remark on the poor results for GRU. 
For the LSTM cell, the PRC error is on average rather good and it clearly shows a dependence on the amount of data provided, approximately at 100 periods worth of data the error saturates around the value 0.1, see Fig.~\ref{fig:datastudy1}(a). 
In the case of GRU the PRC error is large and does not seem to improve with greater amounts of data, Fig.~\ref{fig:datastudy1}(b). 
That is not to say that GRU intrinsically can not perform this task, it might just require a larger network to achieve the same effect - recall that we use the same number of nodes throughout this work. 
GRU was designed as a clever, faster simplification of the LSTM cell. It merges the hidden cell state into the regular cell state as well as merging several logical operations into fewer ones~\cite{understanding_RNN}. 
These simplifications are reasoned by its developers~\cite{GRU} but apparently noticeably impair the cell in performing our particular task. 

The error of the signal undoubtedly should depend on the interval $\Delta$ over which it is evaluated. 
Even with a near perfect model the small errors build up and after a long time the true and reconstructed signals become incoherent, 
which means that with increasing $\Delta$ the error should grow. We see that for both cells in Fig.~\ref{fig:datastudy1}(c,d), although on average the GRU signal errors are significantly larger. 
As with the PRC error, the signal error decreases with the amount of data. 

\begin{figure}[H]
\includegraphics[width=0.95\columnwidth]{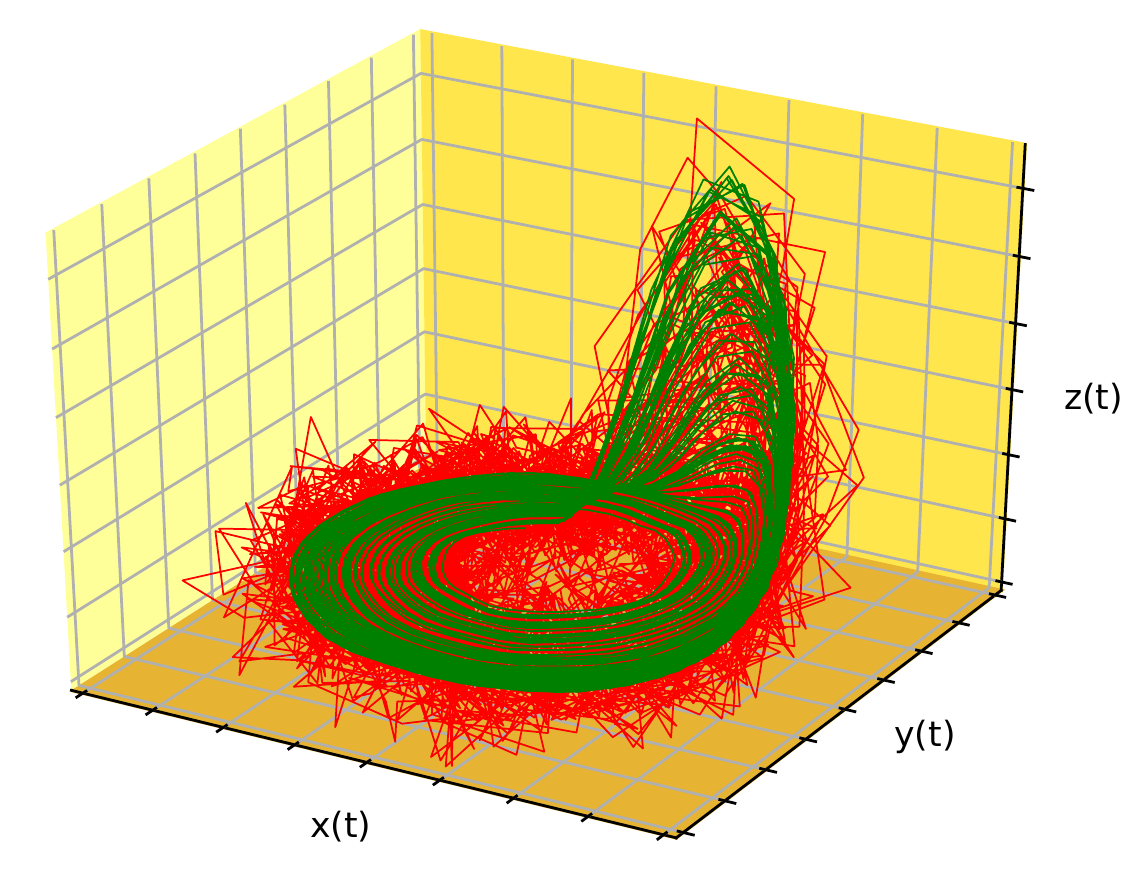}
\caption{The training data in red and the RNN reproduced attractor in green.}
\label{fig6}
\end{figure}

Now for the robustness of  the inference against measurement noise. 
We only present a basic study where we consider the Roessler system, Eq.~(\ref{eq:roessler}) with p-input $p(t) = q(t)$, Eq.~(\ref{eq:inp1}). 
We fix $b = 0.6$ which corresponds to a chaotic regime. Then, we add to each time point a random uncorrelated Gaussian number with mean 0 and standard deviation 1 to represent strong measurement noise, and train the network on the noisy signals. 
We introduce 10000 average periods worth of training data over 500 epochs. 
The network effectively extracts the relevant dynamics and reproduces the attractor well, see Fig.~\ref{fig6}.

\section{\label{sec:methods}Methods}

In this section we specify the methods we used to evaluate the properties of oscillatory systems. For each property we write how we computed it from the equations as well as how we computed it from the RNN.  

\subsection{\label{sec:period_measure}Natural period estimation}
The period is measured as the time between two successive signal-threshold crossings from bellow when the system is unperturbed. 
From equations, the time of crossing is accurately estimated using the H\'enon trick~\cite{Henon-82}. 
When estimating from a network, a linear interpolation from a point before and after the threshold crossing is used.

\subsection{\label{sec:PRC_method}Phase response curve estimation}
Firstly the natural period $T_0$ has to be accurately estimated, see section~\ref{sec:period_measure}. 
Then the system in question is weakly and instantaneously perturbed at particular phases $\varphi^*$, i.e. at times $t^* = \frac{\varphi^*}{2\pi} T_0$ after the beginning of a period, $p(t) = \epsilon \delta(t-t^*)$.
Then the evoked phase shift is evaluated as 
\begin{equation}
Z(\varphi^*) = 2\pi \frac{nT_0 - \sum_{i=1}^n T_i}{\epsilon T_0}
\end{equation}
where $T_1$ is the period in which the perturbation arrives and $T_2,\ T_3,\ T_4,...$ the periods that follow. 
$n$ counts how many periods we wait to evaluate the shift and since we are looking for the asymptotic shift $n$ should be big enough that the PRC does not depend on it, in this paper we used $n = 5$.  

In the case of the network, the time for inputting perturbations is discrete and the best we can do is input perturbation $\epsilon/\Delta t$ where $\Delta t$ is the time increment between two consecutive points in the unrolled RNN. 

\subsection{\label{sec:Lyapunov_method}Maximal Lyapunov exponent estimation}
For computing the exponents from the true system we use the standard technique, since we have the dynamical equations. 

To estimate the exponent from the RNN a different approach is needed. 
Suppose we have access to all the variables of the original system $\vec{x} = (x_1,\ x_1,...,\ x_{n_x})$. In such case the intuitive method can be used: 
\begin{enumerate}
	\item simulate a trajectory $\vec{x}$ for a long time so it settles to the attractor, 
	\item start a new trajectory $\vec{x^\dagger} = \vec{x} + \vec{p}$ with a small arbitrary perturbation $\lVert \vec{p} \rVert = \delta x$ and evolve both for a short time $\delta t$, 
	\item evaluate the deviation $\Delta = \lVert \vec{x^\dagger}-\vec{x} \rVert$, 
	\item renormalize the second trajectory for the deviation to have the same amplitude as the one we started with $\vec{x^\dagger} = \vec{x} + \delta x * (\vec{x^\dagger}-\vec{x})/\lVert \vec{x^\dagger}-\vec{x} \rVert$, but keep the direction of the perturbation the same so that the maximal exponents takes over in the course of several repetitions,
	\item loop to step $3$ and average the quantity $\frac{1}{\delta t} \log(\Delta/\delta x)$ which tends towards the maximal Lyapunov exponent.
\end{enumerate}
Here $\lVert \cdot \rVert$ stands for the $L_2$ norm: $\lVert \vec{v} \rVert = \Big( \sum\limits_i v_i^2 \Big)^{1/2}$.

The more general approach concerns cases where we do not have access to all the variables but only a few, in the extreme case only one $x_1$ - common when dealing with real data. 
In such case the state of the system has to be characterized with several historical values, $\vec{w} = (x_1(t),\ x_1(t-\Delta t),\ x_1(t-2\Delta t),...)$, 
and then the algorithm above can be used as before. 
This is the case in section~\ref{sec:level3}.

\section{\label{sec:discussion}Discussion}
The aim of this study was to test the predictive capacity of recurrent neural network applied to different oscillatory systems. 
One problem common to all oscillators is that the state space collapses to 
a low-dimensional manifold and therefore any reconstruction only allows the prediction on that inertial manifold. 
However, if perturbed we can achieve a much better understanding of the system around its attractor. 
We even can follow and predict a bifurcation outside the range of values which were provided by the data. 
This is a notable fact and it may as well work for other methods, like symbolic regression.

We have applied the method to a range of oscillatory systems, from a time-delay oscillator with a period-2 orbit~(Sec.~\ref{sec:level5}, Fig.~\ref{fig4}), to an excitatory system~(Sec.~\ref{sec:level4}, Fig.~\ref{fig3}), and finally a chaotic attractor~(Sec.~\ref{sec:level3}, Fig.~\ref{fig2}). 
We demonstrate that the trained neural networks can be probed for dynamical responses. As typical characteristics of oscillatory systems we estimated the phase response curve~\cite{PRC_scholarpedia} (PRC), the spiking rate, and the maximal Lyapunov exponent~\cite{Lyapunov_scholarpedia}. 
Other quantities, such as the Floquet exponent~\cite{Floquet}, the amplitude response, the isochronal structure~\cite{isochrones}, synchronization properties~\cite{Pikovsky-Rosenblum-Kurths-01} etc. could be estimated in a similar way. 
We can say that RNNs provide an effective way of estimating oscillatory properties from timeseries, cf.~\cite{Ota-PRL-2009, Imai-2017, Cestnik-Rosenblum-18, Wolf_lyapunov_algorithm, Rosenstein_lyapunov_algorithm}. Our way of applying them to data is novel and should be explored further not only in the context of oscillations. It is, for example, not clear how well RNNs perform for scaling systems like turbulence.  

Since the success of each machine learning method depends on data, we performed a statistical analysis on how the size of the training data set influences the inference. 
The training data required for an effective inference proved to be reasonably small, with only a few 10 periods sufficing for reliably estimating the mentioned dynamical systems quantities. 
We used two popular recurrent network cells in our study: the long short-term memory cell~\cite{LSTM} and the gated recurrent unit~\cite{GRU}. 
The latter proved to be inferior in performing these tasks (at least for the same network size). We also tested the inference with the addition of measurement noise and it proved to be robust, see section~\ref{sec:data_req}.

Along with this publication, we (RC) published a Python software package, {\it OscillatorSnap}~\cite{RCgithub_page}, available on the Python Package Index (PyPI) as: oscillator\_snap.  
It contains most of the examples shown here as well as an array of high level functions for analyzing oscillatory systems, such as, 
a function that computes the phase response curve or the maximal Lyapunov exponent from dynamical equations as well as from a trained RNN model. 

\section{Author contributions statement}
RC did the computational work, MA brought in RNN, both authors wrote the article.

\section{Acknowledgments}
We thank Nicolas Deschle, Bastian Pietras, Thomas Kreuz, as well as the employees of Ambrosys, Markus, Franz, Tino, Maxim, Thomas and Greta 
for useful discussions. 
This work was funded by the European
Union’s Horizon 2020 research and innovation program under
the Marie Skłodowska-Curie Grant Agreement No. 642563
(COSMOS).

\nocite{*}

%

\end{document}